# Switchable Ferroelectric Photovoltaic Effects in Epitaxial Thin Films of *h*-$RFeO_3$ having Narrow Optical Band Gaps

**Hyeon Han[a], Donghoon Kim[a], Ji Hyun Lee[a], Jucheol Park[b], Sang Yeol Nam[b,c], Mingi Choi[d], Kijung Yong[d], and Hyun Myung Jang[a,*]**

[a]Department of Materials Science and Engineering, and Division of Advanced Materials Science, Pohang University of Science and Technology (POSTECH), Pohang 37673, Republic of Korea

[b]Gyeongbuk Science & Technology Promotion Center, Gumi Electronics & Information Technology Research Institute, Gumi 39171, Republic of Korea

[c]Department of Materials Science and Engineering, Kumoh National Institute of Technology, Gumi 39177, Republic of Korea

[d]Department of Chemical Engineering, Pohang University of Science and Technology (POSTECH), Pohang 37673, Republic of Korea

*Corresponding Author, *E-mail address:* hmjang@postech.ac.kr (H. M. Jang)

## ABSTRACT

Ferroelectric photovoltaics (FPVs) have drawn much attention owing to their high stability, environmental safety, anomalously high photovoltages, coupled with reversibly switchable photovoltaic responses. However, FPVs suffer from extremely low photocurrents, which is primarily due to their wide band gaps. Here, we present a new class of FPVs by demonstrating switchable ferroelectric photovoltaic effects using hexagonal ferrite (*h*-$RFeO_3$) thin films having narrow band gaps of ~1.2 eV, where R denotes rare-earth ions. FPVs with narrow band gaps suggests their potential applicability as photovoltaic and optoelectronic devices. The *h*-$RFeO_3$ films further exhibit reasonably large ferroelectric polarizations (4.7~8.5 $\mu C \cdot cm^{-2}$), which possibly reduces a rapid recombination rate of the photo-generated electron-hole pairs. The power conversion efficiency (PCE) of *h*-$RFeO_3$ thin-film devices is sensitive on the magnitude of polarization. In the case of *h*-$TmFeO_3$ (*h*-TFO) thin film, the measured PCE is twice as large as that of the $BiFeO_3$ thin film, a prototypic FPV. We have further shown that the switchable photovoltaic effect dominates over the unswitchable internal field effect arising from the net built-in potential. This work thus demonstrates a new class of FPVs towards high-efficiency solar cell and optoelectronic applications.

## 1. Introduction

Ferroelectric photovoltaics (FPVs) belong to metal-oxide (MO) photovoltaics that are known to be chemically stable and environmentally safe [1]. They can be manufactured inexpensively under ambient conditions. In particular, the FPVs function both as photon absorbers and charge separators. Hence, FPVs can be manufactured as a single active-layered structure. The most outstanding feature of FPVs is that the photovoltage can be a few orders of magnitude larger than the band gap of ferroelectrics due to the bulk photovoltaic effect [2-5]. For example, the measured photovoltage of $BiFeO_3$ thin film is as high as ~200 V when the photocurrent direction is perpendicular to the domain wall [6,7]. Furthermore, FPVs show a reversibly switchable photovoltaic effect by changing the polarization direction with the aid of a bias electric field [8-11].

Until recently, the FPV effect has remained as an academic interest rather than having practical applications owing to extremely low photocurrent densities of FPVs in the order of $nA \cdot cm^{-2}$ ~ $\mu A \cdot cm^{-2}$ [9-13]. The observed very low photocurrent density, which is a main drawback of FPVs, is attributed primarily to wide band-gap ($E_g$) characteristics of typical ferroelectric materials applied to FPV devices: $E_g$ of ~2.7 eV for $BiFeO_3$ (BFO) [10], ~3.6 eV for $Pb(Zr,Ti)O_3$ (PZT) [14], and ~3.5 eV for $BaTiO_3$ [15]. Consequently, extensive studies have been made to reduce $E_g$ by suitable chemical modifications. However, this type of the band-gap tuning usually leads to deterioration of ferroelectric or dielectric properties [16]. Notwithstanding the band-gap problems, the research activity of FPVs has been stimulated by the three recent breakthroughs: (i) achievement of the power conversion efficiency (PCE) of 8.1 % by band-gap tuning of $Bi_2(Fe,Cr)O_6$ ferroelectric multilayers [17], (ii) attainment of the PCE exceeding the Shockley–Queisser limit in a $BaTiO_3$ single crystal [18], and (iii) observation of pronounced switchable photovoltaic effects in organometal trihalide perovskite devices [19-21]. For these reasons, investigations on ferroelectric-based photovoltaics are being actively carried out not only using inorganic oxide materials but also using organic-inorganic hybrid materials [22-32].

Herein, we present switchable photovoltaic effects observed in a new class of FPVs, hexagonal rare-earth ferrite thin films. Therese ferrite materials (*h*-$RFeO_3$; R = Y, Dy-Lu) tend to have a narrow band gap of ~1.2 eV, which is in sharp contrast to typical ferroelectric materials having wide band gaps [10,14,15] and suggests their potential applicability as photovoltaic and optoelectronic devices. We found that the power conversion efficiency (PCE) of *h*-$RFeO_3$ thin-film devices is sensitive on the magnitude of ferroelectric polarization. In the case of *h*-$TmFeO_3$ thin film, the measured PCE is twice as large as that of the $BiFeO_3$ thin film, a prototypic FPV. We have further elucidated that the switchable photovoltaic effect dominates over the unswitchable internal field effect which arises from the net built-in potential developed in the ITO/*h*-$RFeO_3$/Pt heterojunction.

## 2. Experimental Section

### *2.1. Fabrication of Thin-Film Heterojunction Devices*

An epitaxial Pt(111) film adopted as the bottom-electrode layer was grown on the $Al_2O_3$(0001) substrate using RF magnetron sputtering. Pulsed laser deposition (PLD) method was then used for the fabrication of hexagonal $RFeO_3$ (*h*-RFO hereafter) films on the Pt(111)/$Al_2O_3$(0001) substrate at a laser energy density of 1.5 $J \cdot cm^{-2}$ with the repetition rate of 5Hz. The substrate was maintained at 830℃. For the fabrication of solar cells having an ITO/*h*-

RFO/Pt heterojunction structure, transparent ITO top electrodes were deposited by PLD through a shadow mask with circular apertures (100~200 $\mu$m in diameter).

*2.2. Characterizations of Thin-Film Devices*

We have performed structural analysis by X-ray diffraction (XRD) to confirm a hexagonal phase as well as in-plane epitaxy in the PLD grown *h*-RFO film layer. For ferroelectric characterization, *P-E* hysteresis loops with a virtual ground mode were obtained using a Precision LC system (Radiant Technologies, Inc.). Atomic-scale structures of *h*-$LuFeO_3$ and *h*-$TmFeO_3$ (hereafter *h*-LFO and *h*-TFO, respectively) thin films were examined by employing high-resolution transmission electron microscopy method (JEM-ARM200F, JEOL with a Cs-corrector) under 200-kV acceleration voltage. For experimental study of the optical bandgap, optical absorption spectra were recorded as a function of the photon energy using a double-beam UV–Vis–NIR spectrophotometer (JASCOV-570). Ultraviolet photoelectron-spectroscopy (UPS; AXIS Ultra DLD) measurements were used to estimate the work functions, the Fermi energies, and the valence-band edges. UPS measurements were carried out using He I (21.22eV) photon lines from a discharge lamp. X-ray photoelectron spectroscopy was used to measure the O1s signal of *h*-RFO thin films having some oxygen-vacancy defects. The current density–voltage (*J-V*) characteristics were measured using a source meter (Compactstat, IVIUM tech.) under simulated AM 1.5G illumination (100 mW·cm$^{-2}$) provided by a solar simulator (Sun 3000, Abet tech.). The incident light intensity was calibrated with a Si solar cell (as a reference) equipped with an IR-cutoff filter (KG-5, Schott).

## 3. Results and Discussion

### *3.1. Epitaxial Film Growth and Ferroelectricity*

Theta-2theta X-ray diffraction (θ-2θ XRD) patterns show that both PLD-grown *h*-TFO and *h*-LFO films (~250-nm-thin) are highly c-axis oriented (Fig. 1a) on a Pt(111)/$Al_2O_3$(0001) substrate. The phi-scan spectra further reveal that *h*-LFO, *h*-TFO and Pt layers are all grown epitaxially with a six-fold hexagonal symmetry (Fig. 1b). This indicates an absence of in-plane 30°-rotation during the growth often observed in hexagonal thin films [33]. The phi-scan patterns were obtained by keeping the Bragg angle at $(11\bar{2}2)$ for *h*-RFO and $(200)$ for Pt. The hexagonal structure of *h*-RFO having the noncentrosymmetric $P6_3cm$ crystal symmetry is shown in Fig. 1c. This polar structure of *h*-RFO is characterized by (i) the $RO_8$ units having trigonal $D_{3d}$ site symmetry and (ii) the $FeO_5$ bipyramids with the $D_{3h}$ site symmetry [34–37]. The asymmetric vertical shift of rare-earth ion (R) with respect to the two neighboring apical oxygen ions is known to be the origin of the c-axis-oriented hexagonal ferroelectricity in *h*-RFO [34–37].

The room-temperature polarization-electric field (*P-E*) curves (Fig. 1d) demonstrate that the remanent polarization ($P_r$) is ~4.7 μC·cm$^{-2}$ for *h*-LFO film and ~8.5 μC·cm$^{-2}$ for *h*-TFO film. These $P_r$ values are comparable to the previously reported values for *h*-RFO films [36,37]. These polarization values seem to be large enough to separate photo-generated electron-hole pairs in a photovoltaic material. We have further confirmed the epitaxial growth of *h*-LFO film by using high-angle annular dark-field scanning-transmission electron microscopy (HAADF-STEM) method. The HAADF-STEM image (Fig. 1e) shows a characteristic 'up-up-down' rumpling pattern of Lu atoms, which indicates a polar nature of the *h*-LFO film along the c-axis. The corresponding SAED pattern is shown in Fig. 1f. The zone axis [110] is parallel to the corresponding diffraction plane for the XRD, i.e., (00l), which indicates that the *h*-LFO film is aligned along the hexagonal *c*-axis. Exactly the same type of HAADF-STEM image and SAED pattern were observed in the epitaxially grown *h*-TFO film.

### *3.2. Band Gap and Solar Absorption Rate*

Ultraviolet-visible-near infrared absorption spectra of *h*-LFO and *h*-TFO are shown in Fig. 2a as a function of the photon energy. The optical band gap ($E_g$) of *h*-RFO is evaluated by adopting the Tauc plot (Fig. 2b) of the absorption spectra. The Tauc model is represented by $(\alpha E)^{1/n} \propto A(E - E_g)$, where $\alpha$ is the absorption coefficient, $E$ is the photon energy ($h\nu$), $E_g$ is the optical band gap, and $A$ is the photon-energy-dependent constant [38]. The power-law exponent, $n$, depends on the transition type, where $n$ = 1/2 for a direct band-gap-allowed transition and $n$ = 2 for an indirect band-gap-allowed transition. According to the previous experimental studies [39,40], *h*-RFO can be classified as a direct band-gap materials. Currently, two distinct values of $E_g$ are reported for *h*-RFO**:** ~1.1 eV [39] or ~2.0 eV [40,41]. Our optical absorption spectra (Fig. 2a) indicate that $E_g$ at ~1.1 eV is related to a broad weak peak at near-infrared region, whereas $E_g$ at ~2.0 eV corresponds to the onset of strong optical absorption at visible region. Accordingly, we have performed the Tauc plot for a direct band-gap transition ($n$ = 1/2) in the vicinity of two characteristic photon-energies. As shown in Fig. 2b, the optical band gap corresponding to the onset of weak absorption is about 1.2 eV**:** 1.18 eV for *h*-LFO and 1.25 eV for *h*-TFO. On the other hand, the photon energy corresponding to the onset of strong optical absorption is evaluated to be ~2.08 eV**:** 2.07 eV for *h*-LFO and 2.09 eV for *h*-TFO (Fig. 2c).

The computed electronic band structures of $h$-LFO and $h$-TFO are shown in Fig. 2d and 2e, respectively. Herein, we have adopted the GGA+$U_{eff}$ method [42] with the Hubbard $U_{eff}$ of 3.5 eV to evaluate the exchange-correlation functional. The optimal value of $U_{eff}$ is chosen by comparing the computed $E_g$ with the experimental value (~1.2 eV). According to the computed band structures, both $h$-LFO and $h$-TFO reveal a direct band-gap transition at Γ or A point of the Brillouin zone. This prediction is supported by the previous experimental study [39]. The conduction band minimum is represented by Fe $3d$ states while the valence band maximum is described by the hybridization of Fe $3d_{z^2}$ and O $2p_z$ states. Thus, the band gap in $h$-RFO mainly originates from the Fe *d*-*d* transitions [39–41].

In Fig. 2f, we compare the thickness-dependent solar absorption of $h$-LFO and $h$-TFO with that of BFO, a prototypic FPV. Here, the fractional amount of solar absorption is calculated using the solar irradiance and absorption coefficient data given in Fig. 2a. The equation of solar absorption used in our evaluation is given by [43]

$$\text{Solar absorption (\%)} = 1 - \frac{\int_{\lambda_{E_g}}^{\infty} S_\lambda d\lambda}{\int_0^{\infty} S_\lambda d\lambda} - \frac{\int_0^{\lambda_{E_g}} e^{(-\alpha d)} S_\lambda d\lambda}{\int_0^{\infty} S_\lambda d\lambda} \quad (1)$$

where $S_\lambda$ is the spectral distribution of the solar irradiance (in $W \cdot m^{-2} nm^{-1}$), $\lambda$ is the solar wavelength (in *nm*), $\alpha$ is the absorption coefficient (in $nm^{-1}$), and $d$ is the film thickness (in *nm*). At the film thickness of 250 nm, the solar absorption rate of the $h$-LFO film is 1.6 times bigger than that of the BFO film. For the thicknesses greater than 5 μm, the solar absorption rate of the $h$-LFO film shows a plateau behavior and is 3.2 times bigger than that of the BFO film. Since the band gaps of $h$-RFOs (≤~2.0 eV) are significantly narrower than typical FPVs such as BFO and PZT ($E_g$ between 2.7 and 3.6 eV) [10,14], the photon absorption is expected to be much more pronounced in $h$-RFOs, especially in the visible light region (Fig. 2a). Thus, the predicted larger solar absorption (Fig. 2f) can be attributed primarily to the narrower band-gap characteristic of the *h*-RFO in comparison with the BFO film. This is the main advantage of *h*-RFOs as compared with other conventional FPVs. In view of this, *h*-RFOs are expected to be promising materials for future FPV devices.

### *3.3. Ferroelectric Photovoltaic Effects and PCE*

To examine the ferroelectric photovoltaic responses of *h*-RFO films, we have fabricated a solar cell having an ITO/*h*-RFO/Pt heterojunction structure (Fig. 3a), where ITO denotes a transparent indium tin oxide top-electrode layer. Two opposite electrical-poling directions were used to examine the switchable photovoltaic effect**:** "upward poling" signifies the application of a positive voltage to the bottom electrode (Pt), whereas "downward poling" denotes the application of a negative voltage to the bottom electrode. To ensure a complete polarization switching by the poling, we applied an electric field of 1.5 $MV \cdot cm^{-1}$, which is much stronger than the coercive field ($E_c$), ~0.5 $MV \cdot cm^{-1}$ (Fig. 1d). Figure 3b and 3c show illuminated *J-V* characteristics of ITO/*h*-LFO/Pt device and ITO/*h*-TFO/Pt device, respectively. For comparison, *J-V* characteristics for the ITO/BFO/SRO device is also shown in Fig. 3d. Figure 3e compares the *J-V* curves of these three heterojunction devices under the same upward poling, and indicates a remarkably enhanced photocurrent in the ITO/*h*-TFO/Pt device. Figure 3f shows the time-dependent photocurrent under a zero-bias voltage. The ON and OFF states are repeatable and stable, which clearly demonstrates the photo-induced current in the absence of any bias field.

As presented in Table 1, the PCE of the *h*-LFO device under the upward poling is ~0.001 %,

which is comparable with that of the BFO device. In contrast, the PCE of the *h*-TFO device is ~0.002 %, which is twice as large as that of the BFO device. To identify the origin of the enhanced PCE observed in the *h*-TFO device, we have compared $E_g$ and $P_r$ values of these three relevant heterojunction devices as these parameters are known to greatly influence the photovoltaic efficiency of FPVs. The *h*-TFO and *h*-LFO devices show similar values of $E_g$ (Fig. 2b) and the solar absorption rate at the film-thickness of 250 nm (Fig. 2f). In contrast, $P_r$ of the *h*-TFO device is ~1.8 times higher than that of the *h*-LFO device (Fig. 1d)**:** 4.7 $\mu C \cdot cm^{-2}$ for *h*-LFO *versus* 8.5 $\mu C \cdot cm^{-2}$ for *h*-TFO. The enhanced polarization in a film tends to increase the depolarization-field gradient, which promotes an effective separation of the photo-generated electron-hole (*e-h*) pairs. This consequently leads to increased photocurrent density and PCE in the *h*-TFO device, as compared with the *h*-LFO device. In addition, enhanced concentration of oxygen-vacancy defects can modulate the energy band and thus affect the photovoltaic response by reducing the barrier height at the interface [11,44]. This modulation would be possible by the migration of oxygen vacancies to the polarization-head direction during the electrical poling. According to our estimate based on the X-ray photoelectron spectra (XPS), however, the concentration of oxygen vacancies in the *h*-LFO film is effectively equal to that of the *h*-TFO film (Fig. S1). Therefore, the observed enhanced PCE in the *h*-TFO film (Table 1) can be attributed to the increased depolarization-field gradient, rather than to the enhanced concentration of oxygen-vacancy defects.

We further examined the effect of the film thickness on the $J-V$ responses in the *h*-TFO and BFO devices (Fig. 4 and Table 2). Both devices show that $J_{sc}$ increases gradually with decreasing the film thickness. This is primarily owing to the enhanced internal field with decreasing thickness, which is a combined effect of the depolarization field and the Schottky-junction barrier.[8] Notably, as the film thickness increases, the PCE difference between $h$-RFO and BFO also generally increases (Fig. 4c). i.e., as the film thickness increases to 250, 800 and 1500 nm, the PCE ratio of $h$-TFO to BFO also increases to 2.0, 3.5, and 4.6 times, respectively. This is mainly because the absorption amount of the light below ~2 eV in $h$-RFO, which shows low absorption coefficients (Fig. 2a), increases as the film thickness increases to µm level. This can be demonstrated by the thickness dependent solar absorption rate as previously shown in Fig. 2f.

### *3.4. Origin of Asymmetric Switchable Photovoltaic Responses in h-RFO*

Another noticeable feature of the $h$-RFO-based solar cells is that the PCE under the upward poling is substantially higher than that under the downward poling (Table 1). These asymmetric photovoltaic responses can be attributed to (i) the difference in the Schottky barrier height between the top and bottom interfaces (that is, net built-in potential) and (ii) the asymmetric spatial distribution of defects, typically oxygen vacancies [10,44]. To understand the observed asymmetric photovoltaic responses, we have first examined the characteristic energy levels of heterojunction devices (Fig. S2b), which were extracted from the ultraviolet photoelectron spectra (UPS) shown in Fig. S2a. The work function ($\varphi$) is evaluated by using the four $E_{cut\text{-}off}$ values presented in the UPS spectra and by subsequently applying these values to the following equation: $\varphi = 21.22$eV (He I) – $E_{cut\text{-}off}$. The results are: $\varphi = 4.40$ eV for ITO, $\varphi = 5.30$ eV for Pt, and $\varphi(= -E_f) =$ 4.60 eV for $h$-LFO, and 4.58 eV for $h$-TFO. For the $h$-LFO or $h$-TFO layer, $(E_v - E_f)$, thus $E_v$ value, is determined by a linear extrapolation of the low binding-energy region of UPS [45]. Finally, the electron affinity, $E_c$, can be evaluated by using the previously estimated bandgap $(E_c - E_v)$ and $E_v$. Figure S2b graphically summarizes all the estimated characteristic energy levels.

A schematic energy band diagram can be extracted from these estimated characteristic energy levels ($\varphi, E_f, E_v$ & $E_c$ in Fig. S2b). Figure 4a and 4b, respectively, show these diagrams for the ITO/$h$-LFO/Pt and ITO/BFO/Pt heterojunctions. Here, the Schottky barrier height at the Pt/$h$-LFO junction is evaluated by $\Phi_{Pt/LFO} = \varphi_{Pt} - \varphi_{LFO}$ = 5.30−4.60 = +0.70 eV. On the other hand, the Ohmic antibarrier depth at the ITO/$h$-LFO contact is given by $\Phi_{ITO/LFO} = \varphi_{ITO} - \varphi_{LFO}$ = 4.40−4.60 = −0.20 eV (Fig. S2). Then, the barrier-height difference (or net built-in potential) is given by $\Delta E = \Phi_{Pt/LFO} - \Phi_{ITO/LFO}$ = 0.70−(-0.20) = +0.90 eV. It is interesting to notice that the net built-in potential of the ITO/$h$-TFO/Pt heterojunction is also +0.90 eV: ΔE = 0.72−(-0.18) = +0.90 eV (Fig. S3). Because of this energy gradient caused by the nonzero net built-in potential, an internal bias field develops along the cell. Thus, the photo-generated electrons tend to migrate to the ITO/$h$-RFO interface, whereas the photo-generated holes move towards the Pt/$h$-RFO interface.

The electrical poling can substantially alter the energy band diagram of an FPV cell through the poling-induced switching of the depolarization-field direction [8,44,46]. Under the upward poling, the direction of the depolarization field ($\mathbf{E}_{dp}$) is parallel to the direction of the unswitchable net internal bias field ($\mathbf{E}_{bi} = \mathbf{E}_{bi\text{-}bottom} + \mathbf{E}_{bi\text{-}top}$), which results in the enhanced degree of band bending (Fig. 5) and thus increased photovoltaic efficiency under the upward poling. Under the downward poling, on the contrary, the magnitude of the energy gradient becomes smaller due to the significantly reduced net internal electric field ($\mathbf{E}_{net}$). This is because the depolarization field ($\mathbf{E}_{dp}$) is now antiparallel to the unswitchable net built-in field ($\mathbf{E}_{bi}$), yielding a significantly reduced photocurrent density and PCE value under the downward poling. These modulated energy band diagrams account for the observed asymmetric switchable photovoltaic responses (i.e., difference in the PCE between upward and downward poling). The ITO/$h$-TFO/Pt device also shows similar energy band diagrams as depicted in Fig. S3.

### *3.5. Switchable Photovoltaic Effect vs. Built-in Field Effect in h-RFO*

The contribution of the switchable photovoltaic effect to the net photovoltaic response can be qualitatively estimated by the following equation [10]: $V_{sp} = \frac{1}{2}|V_{+} - V_{-}|$, where $V_{sp}$ denotes the switchable open-circuit voltage component mainly arising from the switchable ferroelectric polarization, and $V_{+}$ and $V_{-}$ are the open-circuit voltage obtained after positive and negative polings, respectively. On the other hand, the unswitchable voltage component caused by the built-in internal field ($V_{bi}$) is given by $V_{bi} = \frac{1}{2}|V_{+} + V_{-}|$. $V_{sp}$ and $V_{bi}$ values of the $h$-RFO films can be estimated using these two equations and the photovoltaic data shown in Fig. 3b and 3c. In the case of $h$-LFO film: $V_{sp} = \frac{1}{2}|-0.40 - 0.18| = 0.29\,V$ and $V_{bi} = \frac{1}{2}|-0.40 + 0.18| = 0.11\,V$. Similarly, for the $h$-TFO film: $V_{sp} = \frac{1}{2}|-0.42 - 0.25| = 0.335\,V$ and $V_{bi} = \frac{1}{2}|-0.42 + 0.25| = 0.085\,V$. Thus, for both $h$-RFO-based films, the switchable photovoltaic effect dominates over the unswitchable (nonferroelectric) internal-field effect which mainly stems from the net built-in potential.

In contrast to the asymmetric switchable photovoltaic responses of $h$-RFO devices, the BFO device exhibits a more symmetric response, *e.g.*, $J_{sc}$ and PCE in Table 1. In the case of ITO/BFO/SRO device, the net built-in potential is estimated to be relatively negligible: $\Delta E = \Phi_{bottom}(0.15\text{ eV}) - \Phi_{top}(0.10\text{ eV}) = 0.05\text{ eV} \ll \Delta E = 0.90\text{ eV}$ for $h$-RFO (Fig. 5). Accordingly, it requires much more energy to switch the current direction of the ITO/*h*-RFO/Pt device than that of the ITO/BFO/SRO device. Moreover, the BFO device possesses a significantly larger polarization value than those of the *h*-RFO devices: $P_r \approx 60$ μC·cm$^{-2}$ for [001]-oriented BFO (Fig. S4b) *vs.* $P_r \approx$ 4.7~8.5 μC·cm$^{-2}$ for *h*-RFO. Thus, as depicted in Fig. 5b, the observed nearly symmetric switchable photo-response in the ITO/BFO/SRO device can be attributed to a large depolarization field ($\mathbf{E}_{dp}$) which is suitably combined with a small (unswitchable) built-in field ($\mathbf{E}_{bi}$). Although the ITO/*h*-TFO/Pt device shows an asymmetric photovoltaic response, its PCE (or $J_{sc}$) is remarkably larger than that of the ITO/BFO/SRO device (Table 1). This observation can be interpreted in terms of a significantly reduced band gap in the *h*-RFO FPVs ($E_g$ of ~2.7 eV for BFO *vs.* ~1.2 eV for *h*-RFO). On the other hand, the depolarization-field effect originating from $P_r$ cannot account for this enhanced PCE in *h*-RFO as $P_r$ of *h*-RFO is much smaller than $P_r$ of BFO.

## 4. Conclusions

We have demonstrated a new class of FPVs using heteroepitaxially grown *h*-RFO thin-film heterostructures, where R = Tm and Lu. The *h*-RFO films show narrow band gaps of ~1.2 eV, which indicates a distinct advantage over other typical FPVs having wide band-gap characteristics. In addition, the *h*-RFO films exhibit reasonably large ferroelectric polarizations (4.7~8.5 μC·cm$^{-2}$). This effectively reduces a rapid recombination rate of the photo-generated *e-h* pairs. The PCE of *h*-RFO thin-film devices is sensitive on the magnitude of ferroelectric polarization, suggesting an important role of the depolarization-field gradient in FPV responses. In the case of *h*-$TmFeO_3$ thin film, the measured PCE is twice as large as that of the BFO thin film, a prototypic FPV. We have further elucidated that the switchable photovoltaic effect dominates over the unswitchable internal field effect arising from the net built-in potential. This work opens a new avenue for developing a new ferroelectric material towards high-efficiency solar cell and optoelectronic applications.

## Acknowledgements

This work was supported by the National Research Foundation Grant funded by the Korean Government (MSIP Grant No. 2016R 1D1A1B 03933253) and by Pohang Steel Corporation (POSCO) through the Green Science Program (Project No. 2015Y060 and No. 2016Y038).

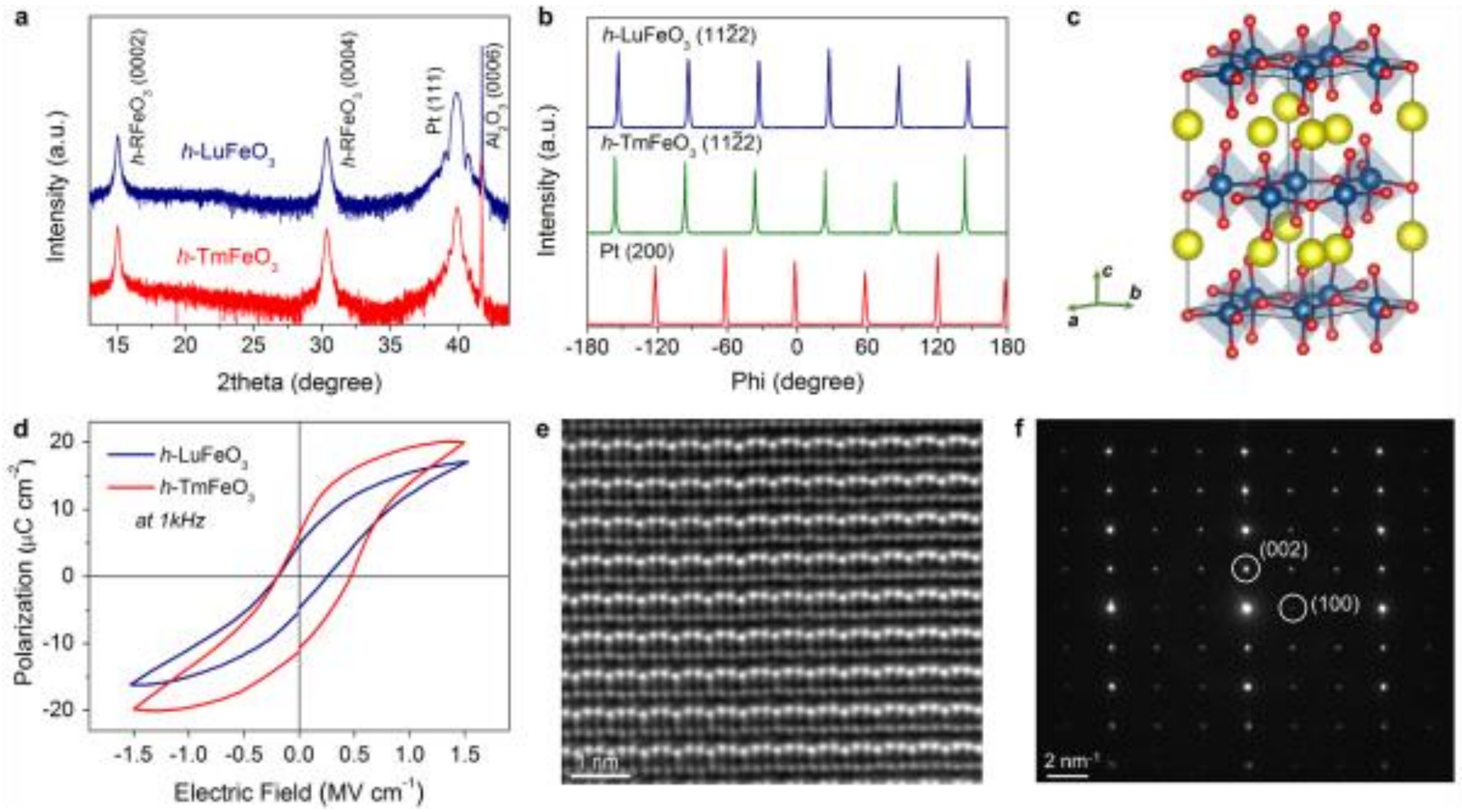

**Fig. 1.** Structural and ferroelectric data of the 250-nm-thick *h*-$RFeO_3$ (*h*-RFO) thin films grown on a Pt(111)/$Al_2O_3$(0001) substrate. (a) Theta−2theta ($\theta-2\theta$) X-ray diffraction (XRD) patterns of the preferential [0001]-oriented *h*-LFO and *h*-TFO films. (b) In-plane XRD phi-scan spectra of *h*-LFO, *h*-TFO, and Pt layers. (c) A schematic crystal structure of the *h*-RFO having the polar $P6_3cm$ symmetry, where dark blue circles denote Fe ions, red circles for oxygen ions, and larger yellow circles designate R (rare-earth) ions. (d) Polarization-electric field (*P*-*E*) hysteresis loops obtained at 300 K using the *ac*-measuring frequency of 1 kHz. (e) A cross-sectional HAADF-STEM image of the *h*-LFO film and (f) the corresponding SAED pattern along the zone axis [110].

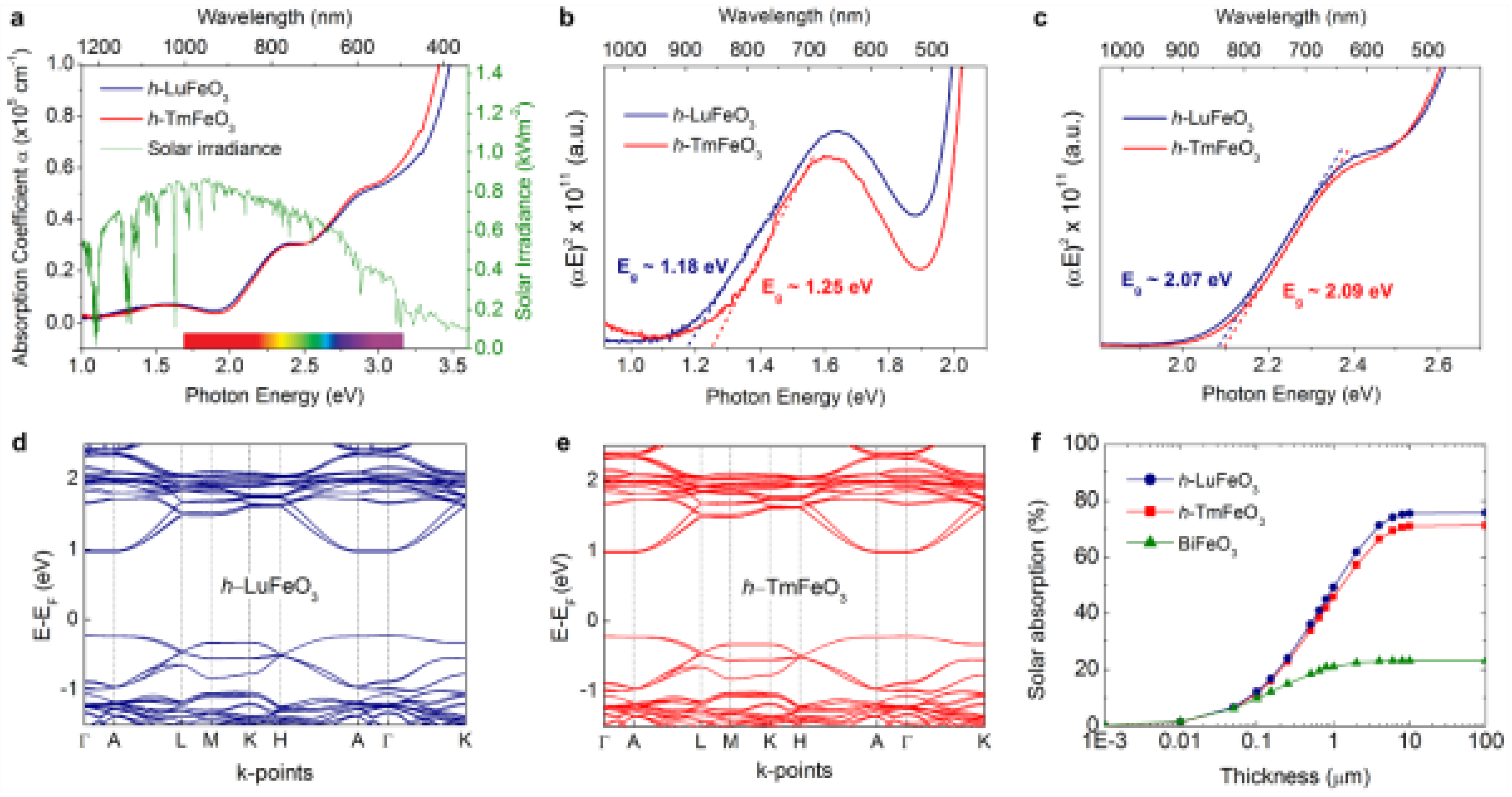

**Fig. 2.** (a) Ultraviolet-visible-near infrared absorption spectra of *h*-LFO and *h*-TFO thin films with the solar irradiance spectrum for the same energy range. Tauc plots of *h*-RFO films near the absorption onset of (b) 1 eV and (c) 2 eV, respectively. Computed band structures of (d) *h*-LFO and (e) *h*-TFO along high-symmetry ***k***-points. (f) Thickness-dependent solar absorption (%) curves for *h*-LFO, *h*-TFO, and BFO films.

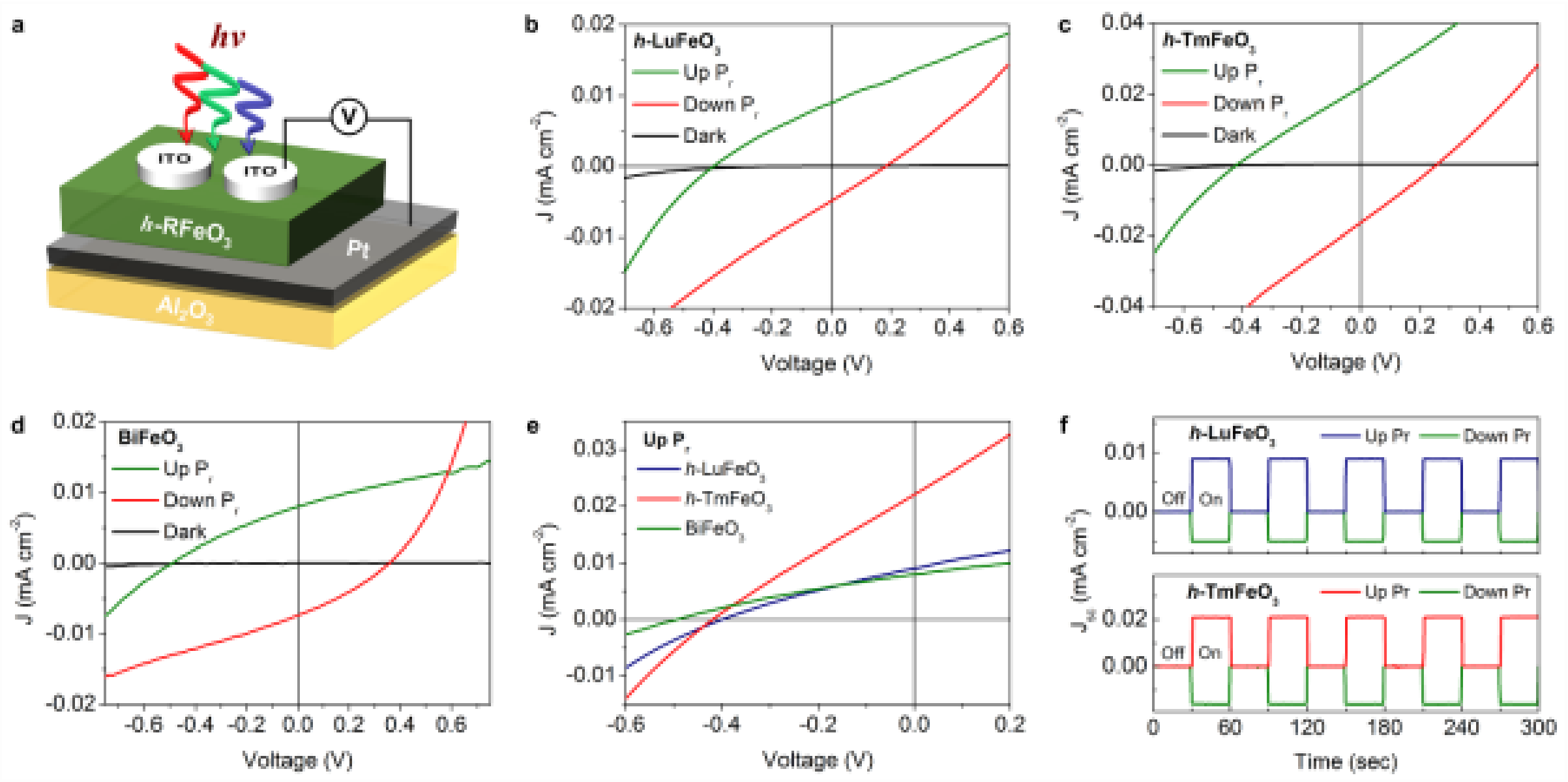

**Fig. 3.** (a) A schematic representation of the ITO/*h*-RFO/Pt heterojunction device. *J*-*V* characteristics of (b) ITO/*h*-LFO/Pt, (c) ITO/*h*-TFO/Pt, and (d) ITO/BFO/SRO devices under AM 1.5G illumination. (e) *J*-*V* curve of the ITO/*h*-TFO/Pt device compared with those of the ITO/*h*-LFO/Pt and ITO/BFO/SRO devices after the upward poling. (f) Zero-bias photocurrent density of the ITO/*h*-LFO/Pt (upper panel) and ITO/*h*-TFO/Pt (lower panel) devices as a function of time.

**Table 1.** Photovoltaic parameters[a] of *h*-TFO/Pt, *h*-LFO/Pt, and BFO/SRO thin-film heterojunction devices (~250-nm-thick) under AM 1.5G illumination.

| Device | Polarization | $J_{SC}$ (mA cm$^{-2}$) | $V_{OC}$ (V) | F.F. (%) | PCE (%) |
|---|---|---|---|---|---|
| *h*-$TmFeO_3$ | Up | 0.021 | -0.42 | 26.2 | 0.0023 |
| | Down | -0.016 | 0.25 | 27.3 | 0.0011 |
| *h*-$LuFeO_3$ | Up | 0.009 | -0.40 | 29.0 | 0.0010 |
| | Down | -0.005 | 0.18 | 25.3 | 0.0002 |
| $BiFeO_3$ | Up | 0.008 | -0.49 | 29.8 | 0.0012 |
| | Down | -0.007 | 0.36 | 31.3 | 0.0008 |

[a]Average photovoltaic efficiencies for 3 to 4 different ITO electrodes.

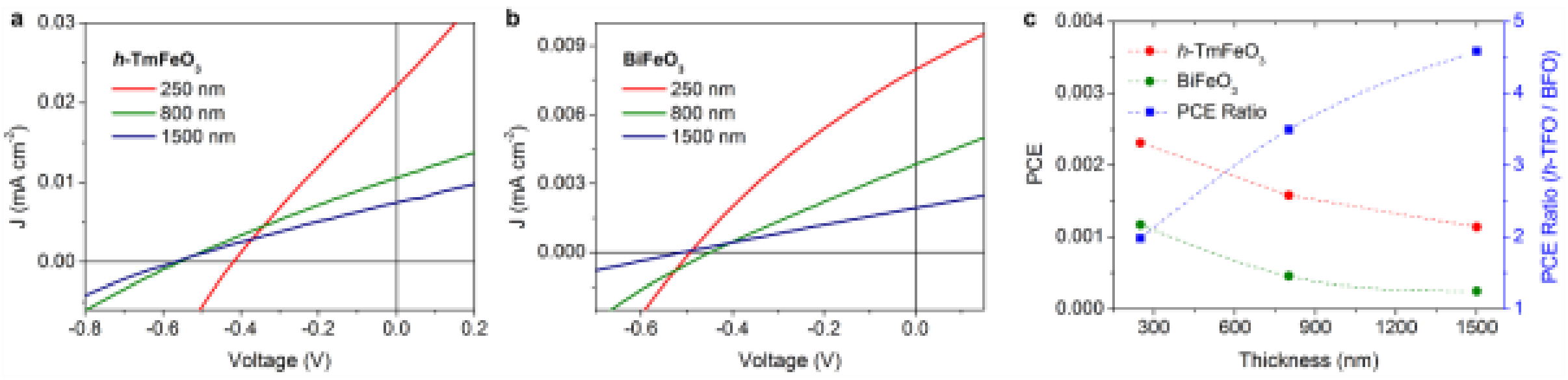

**Fig. 4.** Thickness-dependent $J-V$ characteristics of (a) the ITO/*h*-TFO/Pt devices and (b) the ITO/BFO/SRO devices. (c) PCE of *h*-TFO and BFO devices, and PCE ratio of *h*-TFO to BFO plotted as a function of the film thickness.

**Table 2**. Thickness-dependent photovoltaic parameters[a] of *h*-TFO and BFO Devices under AM 1.5G Illumination.

| Thickness (nm) | Device | $J_{SC}$ (mA cm$^{-2}$) | $V_{OC}$ (V) | F.F. (%) | PCE (%) | PCE Ratio |
|---|---|---|---|---|---|---|
| **250 nm** | *h*-$TmFeO_3$ | 0.0210 | -0.42 | 26.2 | 0.00231 | **2.0** |
| | $BiFeO_3$ | 0.0080 | -0.49 | 29.8 | 0.00117 | |
| **800 nm** | *h*-$TmFeO_3$ | 0.0106 | -0.56 | 26.6 | 0.00158 | **3.5** |
| | $BiFeO_3$ | 0.0038 | -0.45 | 26.4 | 0.00045 | |
| **1500 nm** | *h*-$TmFeO_3$ | 0.0075 | -0.57 | 26.6 | 0.00114 | **4.6** |
| | $BiFeO_3$ | 0.0019 | -0.51 | 25.6 | 0.00025 | |

[a]Average photovoltaic efficiencies for 3 to 4 different ITO electrodes.

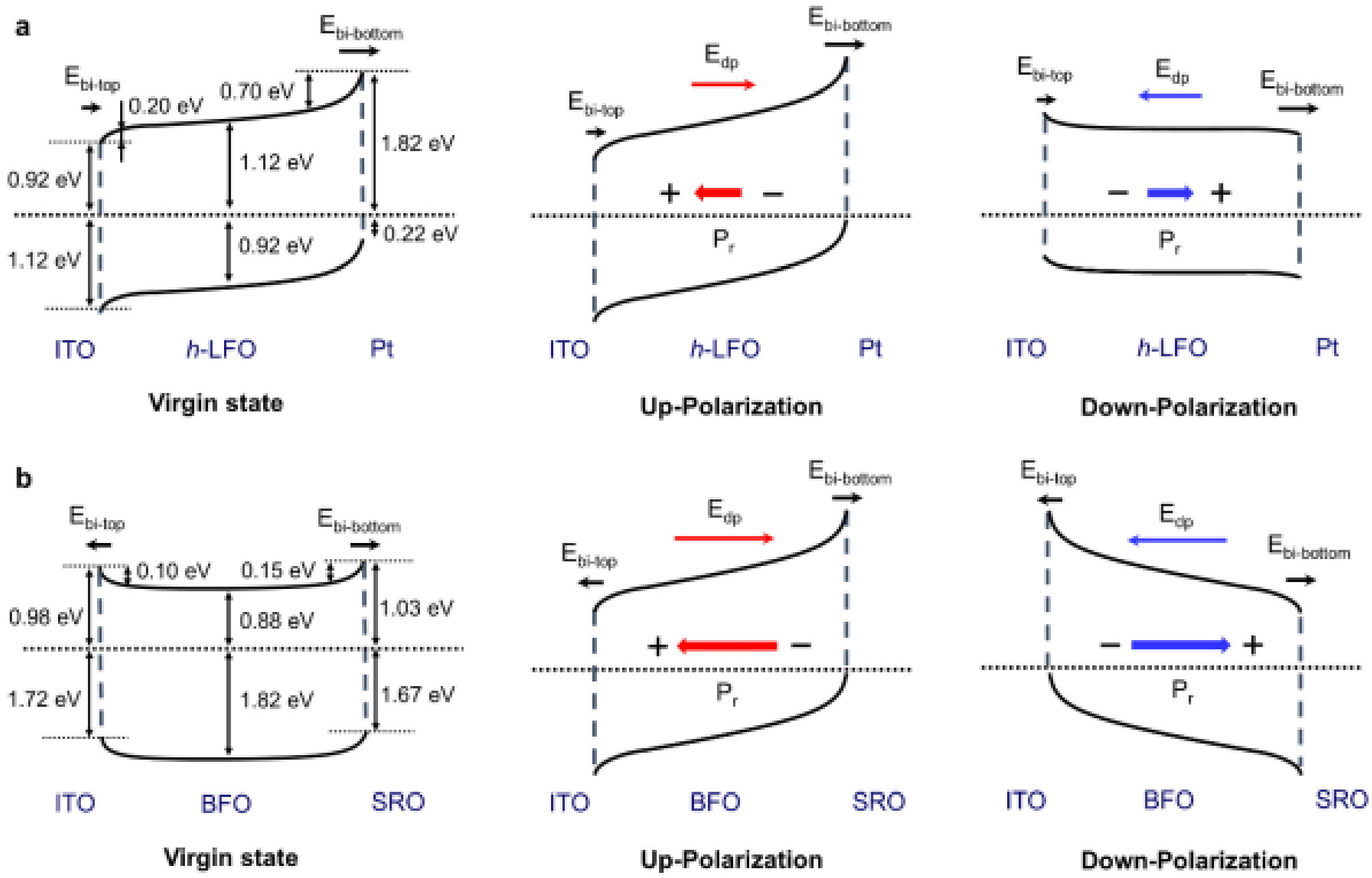

**Fig. 5.** Modulated energy band diagrams of (a) ITO/*h*-LFO/Pt and (b) ITO/BFO/SRO heterojunction devices. Here, the virgin state energy diagram (without poling) is shown on the left-hand side. The up-polarization state diagram (under upward poling) is shown in the middle, whereas the down-polarization state diagram (under downward poling) is shown on the right-hand side.

## Supplementary Information

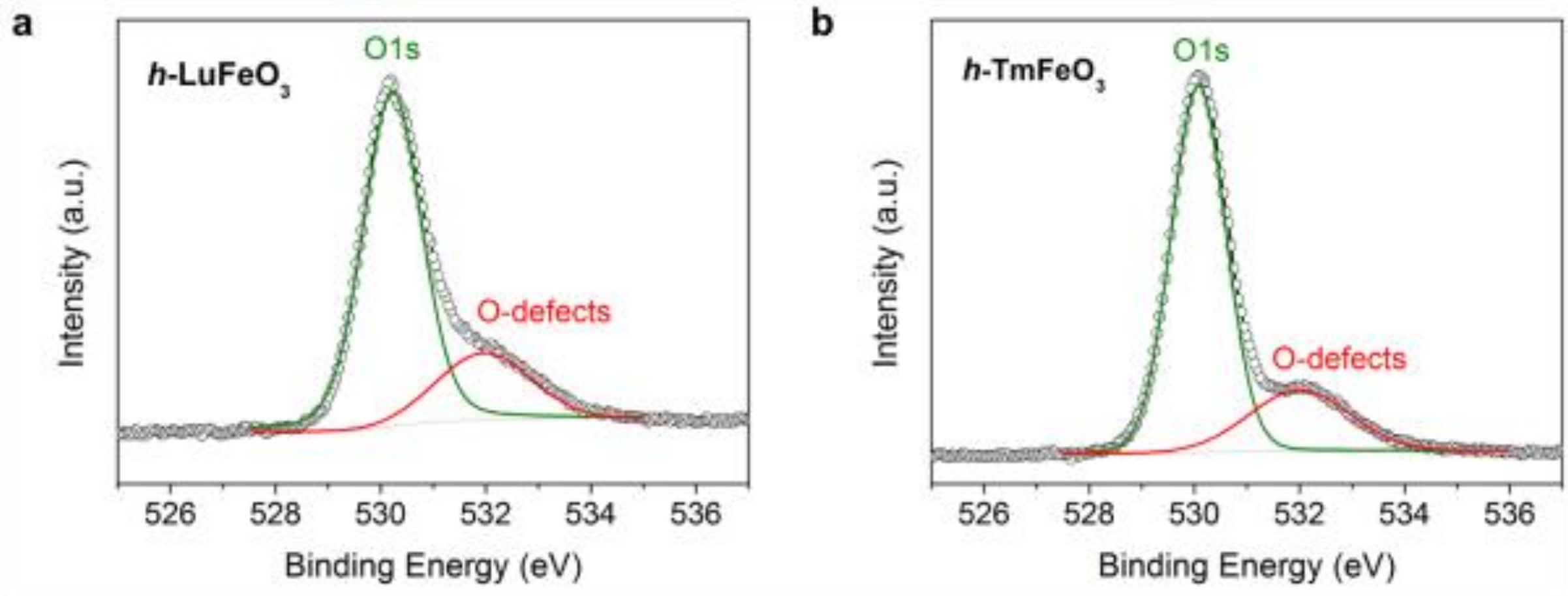

**Fig. S1.** X-ray photoelectron spectra (XPS) of the (a) *h*-LFO and (b) *h*-TFO thin films. The deconvolution of the O1s line results in two peaks of the oxygen (green line) in the *h*-RFO lattice and the oxygen defects (red line).

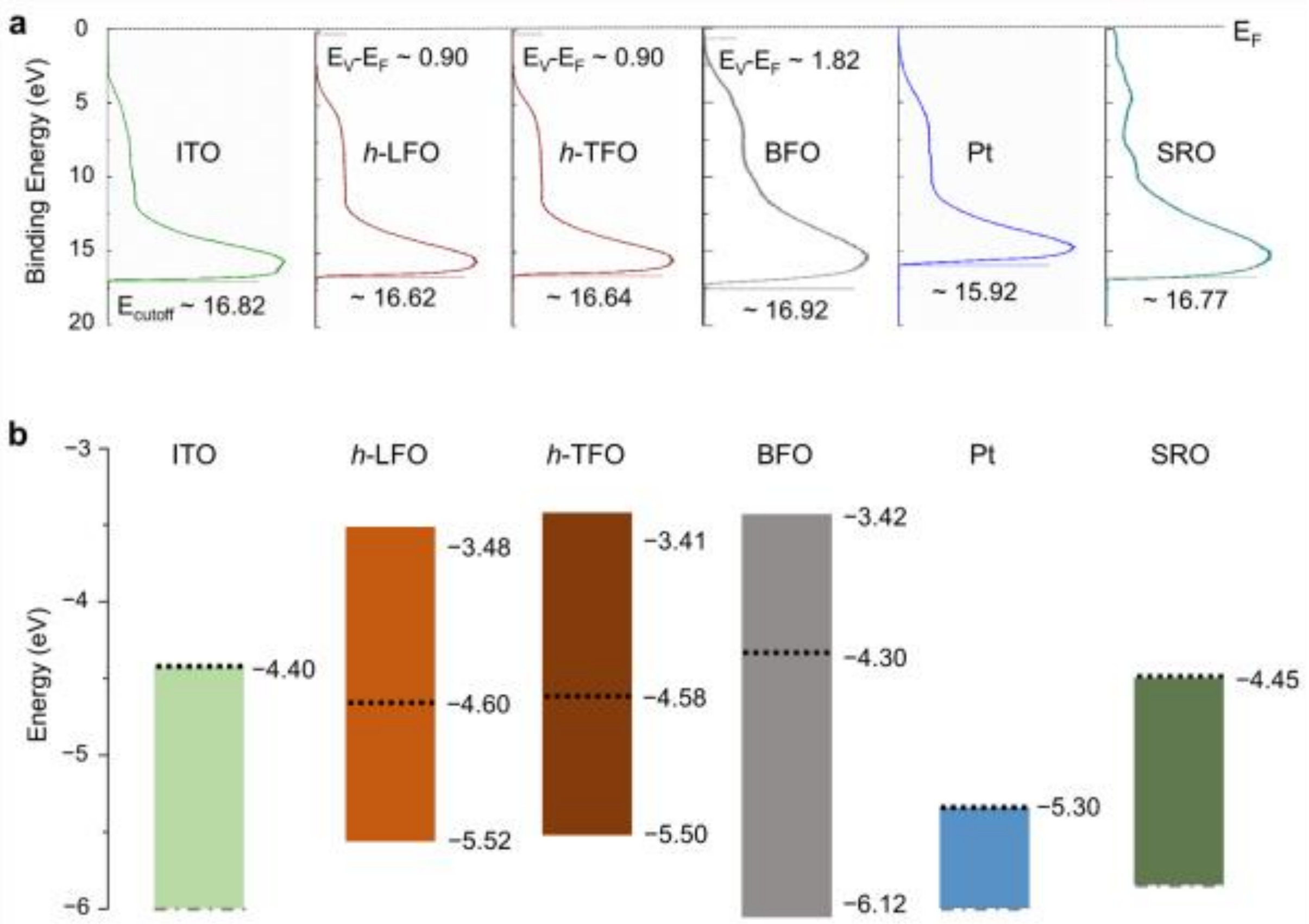

**Fig. S2.** (a) UPS spectra of ITO, *h*-LFO, *h*-TFO, BFO, Pt, and SRO (from the left to the right-hand-side). A low binding-energy region is for the valance-band determination and a high binding-energy region for the work-function determination. (b) An energy level diagram showing the conduction-band minimum, valence-band maximum, and the Fermi level (a dashed line) of each constituting materials.

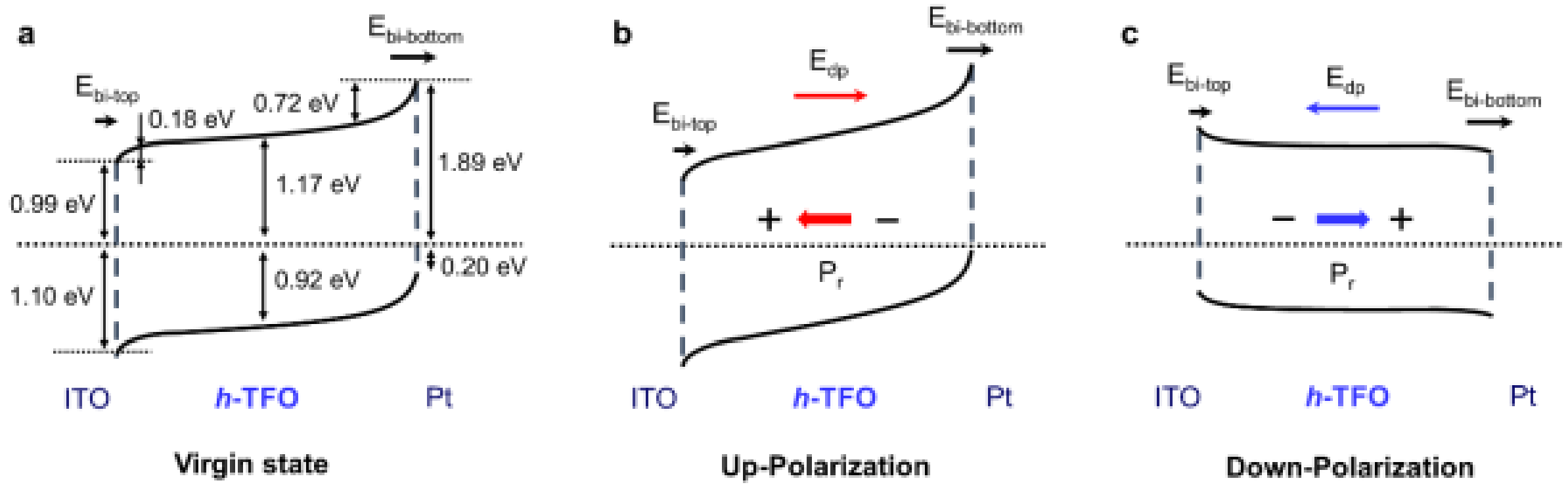

**Fig. S3.** Modulated energy band diagrams of the ITO/*h*-TFO/Pt device. From the left, the virgin state (without poling), the up-polarization state (under upward poling), and the down-polarization state (under downward poling) are shown.

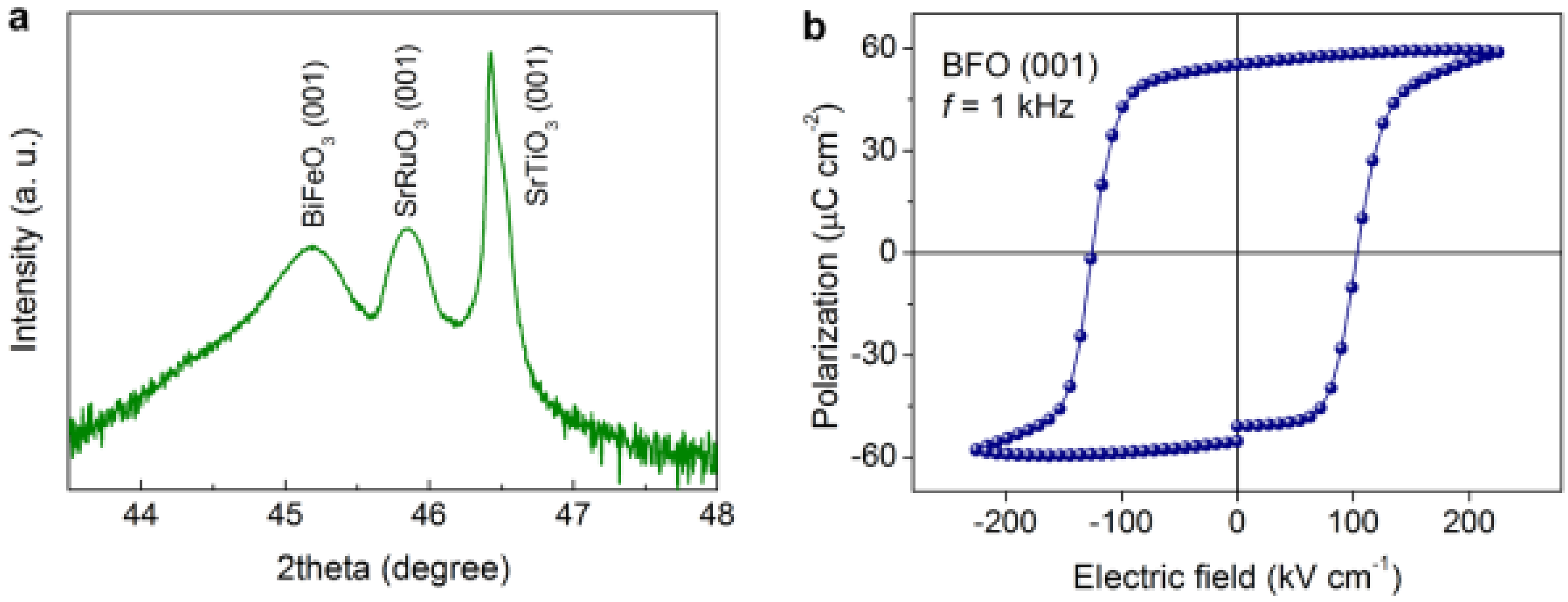

**Fig. S4.** (a) Theta−2theta (*θ−2θ*) X-ray diffraction (XRD) pattern of the preferential [001]-oriented $BiFeO_3$ thin film grown on $SrRuO_3$ (001)/$SrTiO_3$ (001) substrate. (b) A polarization-electric field (*P-E*) hysteresis loop of the 250-nm-thick (001)-oriented $BiFeO_3$ (BFO) layer obtained at 300 K, 1 kHz.